\documentclass[12pt]{article}
\usepackage{latexsym,amsmath,amssymb,fancybox}

\usepackage[english]{babel}

\usepackage{color,epsfig}
\usepackage{graphicx}

\definecolor{navy}{rgb}{0.0,0.0,0.4}

\definecolor{rd}{rgb}{1,0,0}
\definecolor{or}{rgb}{1,.33,0}
\definecolor{pi}{rgb}{.66,.33,.33}
\definecolor{gn}{rgb}{0,.50,0}
\definecolor{be}{rgb}{0,0,.66}
\definecolor{ru}{rgb}{.66,0,.33}
\definecolor{vi}{rgb}{.33,0,.66}
\definecolor{gy}{rgb}{0,.33,.66}
\definecolor{ye}{rgb}{.66,.33,0}
\definecolor{bk}{rgb}{0,0,0}

\def\thf{\baselineskip=\normalbaselineskip\multiply\baselineskip
by 7\divide\baselineskip by 6}

\def\fff{\baselineskip=\normalbaselineskip}

\def\spose#1{\hbox to 0pt{#1\hss}}
\def\lta{\mathrel{\spose{\lower 3pt\hbox
{$\mathchar"218$}}\raise 2.0pt\hbox{$\mathchar"13C$}}}  \def\gta{\mathrel
{\spose{\lower 3pt\hbox{$\mathchar"218$}}\raise 2.0pt\hbox{$\mathchar"13E$}}}

\def\Euro{\spose {\lower 2.5pt\hbox{${^{\bf =}}$}}{ C}}

\def\eqdef{\fff\ \vbox{\hbox{$_{_{\rm def}}$} \hbox{$=$} }\ \thf }

\def\sqr#1#2{{\vcenter{\hrule height.4pt\hbox{\vrule width.8pt height#2pt
\kern#1pt\vrule width.8pt}\hrule height.4pt}}}

\begin{document}

\def\be{\begin{equation}}
\def\fe{\end{equation}}

\newcommand{\eqn}{\label}
\newcommand{\bel}{\begin{equation}\label}

\def\eqdef{\fff\ \vbox{\hbox{$_{_{\rm def}}$} \hbox{$=$} }\ \thf }

\def\ov{\overline}

%SCARLET symbols for physical fields on fibre

\def\Lr{ {\color{rd} {L}} }
\def\Vr{ {\color{rd} {V}} }
\def\Ar{ {\color{rd} {A}} } 

\def\Br{ {\color{rd} {B}} }
\def\Cr{ {\color{rd} {C}} } 
\def\Dr{ {\color{rd} {D}} }

\def\Xr{ {\color{rd} {X}} }\def\Yr{ {\color{rd} {Y}} }
\def\Er{ {\color{rd} {E}} }
\def\Rr{ {\color{rd} {R}} }
\def\nablar{ {\color{rd} {\nabla}} }

\def\calXr{ {\color{rd} {\cal X}} }
\def\calUr{ {\color{rd} {\cal U}} }
\def\calVr{ {\color{rd} {\cal V}} }
\def\calPr{ {\color{rd} {\cal P}} }
\def\calBr{ {\color{rd} {\cal B}} }
\def\Phir{ {\color{rd} {\Phi}} }
\def\Gammar{ {\color{rd} {\Gamma}} }
\def\Thetar{ {\color{rd} {\Theta}} }

\def\dr{\spose {\raise 4.0pt \hbox{\color{rd}{\,\bf-}}} {\rm d}}

%RUBY symbols for gauge fields

\def\Gr{ {\color{ru} {G}} }
\def\Jr{ {\color{ru} {J}} }
\def\calAr{ {\color{ru} {\cal A}} }
\def\calGr{ {\color{ru} {\cal G}} }
\def\calPr{ {\color{ru} {\cal P}} }
\def\calCr{ {\color{ru} {\cal C}} }
\def\ConStruc{ {\color{ru} {\copyright}} }
\def\omegaru{ {\color{ru} \omega}}
\def\Omegaru{ {\color{ru} \Omega}}

\def\alpharu{ {\color{ru} \alpha}}
\def\betaru{ {\color{ru} \beta}}
\def\gammaru{ {\color{ru} \gamma}}
\def\Dru{ {\color{ru} {D}} }
\def\aru{ {\color{ru} {a}} }
\def\Aru{ {\color{ru} {A}} }
\def\Fru{ {\color{ru} F} }
\def\amr{ {\color{ru}\bf{a}} }
\def\Amr{ {\color{ru}\bf{A}} }
\def\Fmr{ {\color{ru}\bf{F}} }
\def\wru{ {\color{ru} {\vert\!\!\vert\!\!\vert}} }

%BLUE symbols
\def\gb{{\color{be} g }}
\def\vb{{\color{be} v }}
%{{\color{be} d }}
%\def\Db{{\color{be} D }} 
\def\nablab{ {\color{be} \nabla}}
\def\Gammab{ {\color{be} \Gamma}}
\def\Thetab{ {\color{be} {\Theta}} }
\def\Ab{{\color{be} A }}
\def\Rb{{\color{be} R}}
\def\db{\spose {\raise 4.0pt \hbox{\color{be}{\,\bf-}}} {\rm d}}

\begin{center}
{\color{rd}\bf  FIELDS IN NONAFFINE BUNDLES. I. \\[0.4cm]
The general bitensorially covariant differentiation procedure.}
\\[1cm]
 \underline{Brandon Carter} \\[0.6cm]
 \textcolor{ru}{Group d'Astrophysique Relativiste (CNRS), 
  Observatoire Paris - Meudon. }
  \\[0.5cm]
 {\color{be} 7 August, 1985.}
\\ [0.5 cm] [Colored version of article in {\it Phys.Rev.}  
{\bf D33} (1986) 983-990].
\\[1.2cm]
\end{center}  .

{\bf Abstract}. The standard covariant differentiation procedure for fields
in vector bundles is generalised so as to be applicable to fields in general
nonaffine bundles in which the fibres may have an arbitrary nonlinear 
structure. In addition to the usual requirement that the base space should 
be flat or endowed with its own linear connection $\Gammab$, and that there 
should be an ordinary gauge connection $\Amr$ on the bundle, it is necessary 
to require also that there should be an intrinsic, bundle-group invariant
connection $\hat\Gammar$ on the fibre space. The procedure is based on the 
use of an appropriate primary-field (i.e. section) independent connector 
$\omegaru$ that is constructed in terms of the natural fibre-tangent-vector
realisation of the gauge connection $\Amr$. The application to 
gauged harmonic mappings will be described in a following article.

\section{Introduction} 
\label{Sec1}

Since at least the time of Clerk-Maxwell, or even earlier, nearly all the
most successsful physical models for the description of the physical world
at a fundamental (and also often at a higher) level have been essentially
based on the conceptual framework of local {\it field theory}. The fields
in question, whose behaviour is governed by local differential equations
of usually not higher than second order, are generally interpretable -- at a
classical level -- as sections of fibre bundles over some appropriate base 
space (which might, for example represent ordinary four-dimensional space-time,
or some higher-dimensional extension or lower-dimensional subspace therof).

In the most familiar and well developed examples (including Yang-Mills theory),
although the {\it theories} themselves may be {\it nonlinear} (in the sense 
that the field equations contain coupling terms of quadratic or higher order)
the actual  {\it fields} are {\it intrinsically linear} in so much as they 
belong to bundles whose fibres are {\it flat}. In the simplest cases the fibre 
space is actually {\it vectorial}, and even in the case of gauge-connection 
fields (e.g. of Yang-Mills type) the fibre space still has a well defined
{\it affine} structure, although there is no longer any preferred origin.
For fields in such essentially linear (i.e. affinely fibreed) bundles, the
standard procedure for the construction of the relevant gauge-covariant
derivatives (in terms of which the field equations are expressed) provided an
appropriate connection is available, is widely known and familiar (see e.g.
Choquet Bruhat, Morette-DeWitt, Bleck-Dillard \cite{1}).

The main purpose of the present work is to describe how the standard machinery 
for gauge covariant differentiation can be generalised so as to be applicable 
to fields that are {\it intrinsically} nonlinear, in the sense of being 
sections of {\it nonaffinely} fibered bubdles. Such nonaffine fields (as 
exemplified by nonlinear $\sigma$ models) have attracted an increasing amount
of interest in recent years.

The usual procedure for ordinary vector bundles needs the provision only of a
{\it gauge connection} $\Amr$, in addition to thee requirement that the 
{\it base space} should either be flat or at least provided with an {\it 
ordinary linear connection} $\Gammab$. The natural generalization to be 
described here requires also that the (curved) {\it fibre space should be
provided with its own linear connection} $\hat\Gammar$.

In a following article we shall describe the application of the general 
purpose formalism set up below to the particular case of a Riemannian
connection induced automatically by the Lagrangian for the natural minimally
gauge coupled generalisation of the class of harmonic mappings that was
described by Misner \cite{2}. These gauged-harmonic mappings will include 
as a special case the gauge-coupled generalization of the nonlinear
$\sigma$ model with fully homogeneous symmetric fibres that was recently
described by the present author \cite{3}.

\section{The concepts of bitensorial differentiation and connector 
fields} 
\label{Sec2}

One of the essential guidelines whose observance qualifies a theoretical
treatment for description as {\it geometric} is the requirement that one
should work as far as possible in terms of entities that are {\it invariant}
in the sense of being independent of any arbitrarily chosen system of 
reference that one might wish to introduce for the sake of explicitness at
some inremediate stage in the treatment. However, the strictest observance 
of this precept risks giving a treatment that either needs to be unduly 
abstract as the price of being elegant or else that needs unweildy 
mathematical machinery as the price of being concrete. For this reason most
theoretical physicists do not insist on the exclusive use of entities that 
are strictly invariant, but as a compromise prefer nevertheless to work 
as far as possible with entities that are at least {\it covariant} in the 
sense of being subject to simply described rules of variation when the 
relevant reference system is altered. One of the simplest and most 
convenient examples is that of quantities represented in terms of sets of 
components whose rules of variation are of {\it tensorial} type in the sense
of being expressible in terms of appropriate contractions with relevant
coordinate transformation matrices. In the specific context of general field
theories we shall be particularly concerned with entities whose covariance
is of {\it bitensorial} in so much as they involve two independent matrices
expressing independent coordinate changes on the base and fibre spaces 
respectively.

As a basic starting point let us consider the case of field $\Vr$ of simple
vectorial type, meaning that its components $\Vr^{_\Ar}$ undergo a change of
the form
{\be \Vr^{_\Ar}\ \mapsto \ \Gr^{_\Ar}_{\ _\Br} \Vr^{_\Br} \label {2.1}\fe}
under the effect of a fibre-coordinate transformation characterised by
the matrix $G^{_\Ar}_{\ _\Br}$. Suppose that we simultaneously carry out
a coordinate transformation
{\be x^\mu \mapsto y^\mu\{x\} \label{2.2} \fe}
on the base space ${\cal M}$ over which the field $\Vr$ is defined,
thereby determining a corresponding base-space transformation matrix
given by
{\be Q^\mu_{\ \nu}=\partial_\nu y^\mu \, ,\label{2.3}\fe}
where we have introduced the abbreviation
$$ \partial_\mu =\frac{\partial}{\partial x^\mu} $$
for partial coordinate differentiation of a field over the base space.
Then the components
{\be \Dru_\mu \Vr^{_\Ar}=\partial_\mu \Vr^{_\Ar}+\omegaru_{\mu\ _\Br}^{\ _\Ar}
\Vr^{_\Br} \label{2.4}\fe}
will qualify for description as those of a {\it covariant} or more
explicitly {\it bitensorial} derivative if they transform according to
the corresponding matrix contraction rule as expressed by
{\be \Dru_\mu \Vr^{_\Ar}\mapsto Q^{-1\nu}_{\, \ \ \ \mu}\Gr^{_\Ar}_{\ _\Br} 
\, \Dru_\nu \Vr^{_\Br} \label{2.5}\, .\fe}

It is evident that the bitensorial covariance property (\ref{2.5}) will
hold if and only if the components $\omegaru_{\mu\ _\Br}^{\ _\Ar}$ have a
covariance property of a rather more complicated nature, namely
{\be \omegaru_{\mu\ _\Br}^{\ _\Ar}\ \mapsto \  Q^{-1\nu}_{\, \ \ \ \mu}
\Gr^{-1_\Cr}_{\, \ \ \ _\Br}(\Gr^{_\Ar}_{\ _\Dr}\omegaru_{\nu\ _\Dr}^{\ _\Cr}
-\partial_\nu \Gr^{_\Ar}_{\ _\Cr}) \, .\label{2.6}\fe}
This  will be of bitensorial form only if the base gradient 
$\partial_\mu \Gr^{_\Ar}_{\ _\Br}$ of the the fibre-coordinate 
transformation matrix $\Gr^{_\Ar}_{\ _\Br}$ happens to vanish (which will
not in general be the case for the examples we wish to consider).

We shall use the term {\it connector} to denote any field
$\omegaru$ having components $\omegaru_{\mu\ _\Br}^{\ _\Ar}$ specified by one
(covariant) base-coordinate index and two (mixed) fibre-coordinate
indices and transforming according to the rule (\ref{2.6}).
A connector can be considered as a special kind of {\it biaffinator},
using the term {\it affinator}  as an abbreviation for affine tensor,
to denote quantities whose components transform according to a rule that
generalises the ordinary kind of tensorial transformation law by 
allowing for the presence of an inhomogeneous additive term [having the
form $-Q^{-1\nu}_{\, \ \ \ \mu}\Gr^{-1_\Cr}_{\, \ \ \ _\Br}\,
\partial_\nu \Gr^{_\Ar}_{\ _\Cr}$ in the example (\ref{2.6})] over and above 
the usual homogeneous multiplicative term [having the form 
$Q^{-1\nu}_{\, \ \ \ \mu} \Gr^{-1_\Cr}_{\, \ \ \ _\Br}\Gr^{_\Ar}_{\ _\Dr}
\omegaru_{\nu\ _\Dr}^{\ _\Cr}$ in the example  (\ref{2.6})].

Insomuch as it is subject to the bi-aftensorial transformation (\ref{2.6}),
a connector $\omegaru$ can be interpreted as a {\it genuine field} in the
sense that it is a {\it section} in an appropriately constructed fibre
bundle $\calCr^\prime$ over the base space ${\cal M}$, the bundle being 
of {\it affine} (rather than ordinary vectorial) kind  in the sense that 
(as well as being subject to the usual group of homogeneous base-coordinate
transformations specified by the matrices $Q^\nu_{\ \mu}$) the fibres of the
bundle are subject to an action of the associated {\it inhomogeneous 
adjoint} group $\calGr^{\prime\dagger}$ of linear transformations 
generated by uniform translations and by the adjoint action of the
matrices $\Gr^{_\Ar}_{\ _\Br}$.

We use the term connector (as distinct from connection) for the purpose of
emphasizing this interpretation of $\omegaru$ as a genuine (biaffinitorial) 
field in the sense of being a section in the relevant (affine) fibre bundle 
$\calCr^\prime$, as chararacterized by an action of the corresponding
inhomogeneous adjoint group $\calGr^{\prime\dagger}$. Of course, such an
$\omegaru$ can also be given a more traditional mathematical interpretation 
as a {\it connection}, meaning an algebra-valued form on an appropriate
{\it principal} fibre bundle $\calPr^\prime$ (see e.g. Choque-Bruhat
{\it et al.}\cite{1} or Carter \cite{4}) associated with the corresponding
vector bundle $\calVr^\prime$ containing $\Vr$, as characterised by the left 
action on itself of the subgroup $\calGr^\prime$ of $\calGr^{\prime\dagger}$ 
generated directly by the multiplicative action of the allowed 
transformation matrices $\Gr^{_\Ar}_{\ _\Br}$. 

The need for rather more care that usual in the interpretation of $\omegaru$ 
-- either as a connector {\it in} $\,\calCr^\prime$ or as a connection 
{\it on} $\,\calPr^\prime$ -- arises in situations where our primary purpose
is to deal with differentiation of a primary field $\Phir$ having values in
a {\it nonaffinely} fibered bundle $\calBr$ subject to the provision of an
ordinary gauge field $\Amr$ with respect to the bundle group $\calGr$ of
$\calBr$. Such a gauge field $\Amr$ will be interpretable in the traditional
way as a connection on the directly associated principle bundle $\calPr$
of $\calBr$ (with nonlinear fibres having the form of $\calGr$ itself) and
it will also be interpretable as a connector field in an appropriate 
affine bundle $\calCr$ subject to the action of the inhomogeneous 
adjoint group $\calGr^\dagger$ associated with $\calGr$ (as well as base
coordinate transformations) on the fibres. 

This {\it primary} connector bundle $\calCr$, containing the gauge section 
$\Amr$ will in the general case (for a nonlinearly fibered primary bundle 
$\calBr$) be distinct from what we shall refer to as the {\it derived} 
principle bundle $\calPr^\prime$ and the derived connector $\omegaru$ (for 
which the corresponding groups $\calGr^\prime$ and $\calGr^{\prime\dagger}$
may be larger than $\calGr$ and $\calGr^\dagger$). These derived bundles and 
the connector $\omegaru$ are not (in the nonlinear case) determined in 
advance by the corresponding primary bundles and the gauge field $\Amr$, but
are specified as functions of the section $\Phir$ in $\calBr$. Any such
section immediately determines a corresponding bundle $\calVr^\prime$ of
ordinary vectorial type (over the same base ${\cal M}$ whose elements
$\Vr$ are just the tangent vectors to the fibres of $\calBr$ at the section
$\Phir$. This section dependent vector bundle $\calVr^\prime$ is the basic
building block from which, in conjunction with the ordinary cotangeent
bundle over ${\cal M}$, one can proceed to construct the corresponding
tensorially associated vector bundles that are needed to contain 
bitensorial derivatives of various orders. The derived bundles
$\calPr^\prime$ or $\calCr^\prime$ that are needed for the definition -- as,
respectively, a connection or a section -- of the connector $\omegaru$
that will be required (for the explicit construction of such
bitensorial covariant derivatives) will be, respectively, the directly 
associated principle bundle $\calPr^\prime$ of $\calVr^\prime$ or the
corresponding affine bundle $\calCr^\prime$ as characterized by the bundle 
group $\calGr^\prime$ of $\calVr^\prime$ and of its (inhomogeneous adjoint) 
extension $\calGr^{\prime\dagger}$ acting on $\calCr^\prime$.

The possibility that the derived bundle group  $\calGr^\prime$ may be
considerably larger than the primary bundle group  $\calGr$ results from 
the fact that it arises from (in general, base -position dependent) fibre
coordinate transformations
{\be \Xr^{_\Ar}\{\Xr,x\} \ \mapsto \ \Gr^{_\Ar}\{\Xr,x\} \label{2.7}\fe}
for $\Xr \in \calXr$, $x\in {\cal M}$, where ${\cal M}$ is the base space
and $\calXr$ the fibre space of $\calBr$, that arise {\it not only} from
the action of the primary gauge group  $\calGr$ but also from the group
of non linear transformations between coordinates of the different patches
that may be needed to cover the fibre space $\calXr$ when it has itself a 
nonlinear manifold structure. In terms of the original fibre coordinates
$\Xr^{_\Ar}$, the elements of  $\calGr^\prime$ will be represented by
matrices of the form
{\be \Gr^{_\Ar}_{\ _\Br}\{\Xr,x\} =  \Gr^{_\Ar}_{\ ,_\Br} \label{2.8}\fe}
as evaluated on the chosed section
{\be \Xr =\Phir\{x\}\, ,\label{2.9}\fe}
where a comma denotes partial differentiation, so that, in particular, the
total space gradient components (with respect to the local coordinates 
$x^\mu$ and $\Xr^{_\Ar}$) that appear in the connector transformation
formula (\ref{2.6}) will be given explicitly by
{\be \partial_\mu \Gr^{_\Ar}_{\ _\Br}= \Gr^{_\Ar}_{\ ,_\Br ,\mu}
+  \Gr^{_\Ar}_{\ ,_\Br ,_\Cr} \Phir^{_\Cr}_{\ ,\mu} \ ,\label {2.10}\fe}
where
$$  \Phir^{_\Cr}=\Xr^{_\Cr}\{\Phir\{x\}\} \, .$$

In the following sections we shall describe the natural procedure for
explicitly constructing a well defined section dependent connector field 
$\omegaru$ obeying the rule (\ref{2.7}), in terms of a previously given 
primary gauge field $\Amr$ and of ordinary linear connections $\Gammab$ and 
$\hat\Gammar$ on the base and fibre spaces ${\cal M}$ and $\calXr$ 
respectively. Before doing so we remark that because such a 
section-dependent connector can be interpreted as as an ordinary connection
on the artificially constructed (section dependent) vector bundle
$\calVr^\prime$, it follows that $\omegaru$ will automatically have the
usual properties that are familiar from the standard theory of fixed
(section-independent) connections. In particular, the connector $\omegaru$
will determine a corresponding well-defined (but section dependent)
{\it bitensorial} curvature field $\Omegaru$  according
to a formula of the familiar form
{\be \Omegaru_{\mu\nu\ _\Br}^{\,\ \ _\Ar}=2\partial_{[\mu}\,
\omegaru_{\nu]\ _\Br}^{\,\ _\Ar}
+2 \omegaru_{[\mu\ |_\Cr|}^{\,\ _\Ar}\omegaru_{\nu]\ _\Br}^{\ _\Cr}
\label{2.11} \fe}
(where square brackets denote antisymmetrisation) and this field will
satisfy a Bianchi identity of the familiar form
{\be \partial_{[\mu}\Omegaru_{\nu\rho]\ _\Br}^{\ \ \ _\Ar}=
\Omegaru_{[\mu\nu\ |_\Cr|}^{\ \ \ _\Ar}\omegaru_{\rho]\ _\Br}^{\,\ _\Cr}
-\omegaru_{[\mu\  |_\Cr|}^{\ \ _\Ar}\Omegaru_{\nu\rho]\,\ _\Br}^{\ \ \ _\Cr}
\, .\label{2.12} \fe}

\section{Bitensorial differentiation in the absence of a gauge 
transformation}
\label{Sec3}

Before dealing with the general situation (where there is a non-trivial
gauge group $\calGr$) let us start by dealing with the comparitively
simple case for which the fundamental bundle $\calBr$ under consideration
is endowed with a trivial {\it direct product} structure $\calXr \times
{\cal M}$ where ${\cal M}$ is the base space, with local coordinates
$x^\mu$, and $\calXr$ is the fibre space, with local coordinates
$\Xr^{_\Ar}$. The imposition of such a diresct product structure is
equivalent to the specification of an {\it integrable connection} on the
bundle. Its presence enables us to restrict our attention for the time 
being to fibre-coordinate transformations
{\be \Xr^{_\Ar}\{\Xr\}\ \mapsto \ \Yr^{_\Ar}\{\Xr\} \label{3.1}\fe}
that are {\it independent} of base position, i.e. such that
{\be \Yr^{_\Ar}_{\ ,\mu}= 0 \, ,\label{3.2}\fe}
unlike the more general transformations of the form (\ref{2.7}) that
were mentioned in the introduction and to which we shalll return in the
next section.

In such integrable cases the procedure described by Misner\cite{2} for
the Riemannian case can be taken over directly provided that the base
${\cal M}$ and the fibre $\calXr$ each has its own linear connection.
An ordinary linear connection on ${\cal M}$ will be specified by a
corresponding purely {\it affinitorial} (as opposed to the more general 
biaffinitorial) connector field $\Gammab$ with mixed components
$\Gammab_{\!\mu \ \rho}^{\ \nu}$ which can be used, e.g. for a simple
tangent vector $\vb$ with components $\vb^\mu$, to specify the covariant
variation $\db\vb$ with components $(\db \vb)^\mu$ associated with an 
infinitesimal component variation ${\rm d}(\vb^\mu)$ in conjunction
with a base displacement ${\rm d}x^\mu$ by the formula
{\be \db \vb^\mu={\rm d}(\vb^\mu)+\Gammab_{\!\nu \ \rho}^{\ \mu}\,\vb^\rho\,
{\rm d}x^\nu \label{3.3}\fe}
so that if $\vb$ is defined as a field over ${\cal M}$ there will be a 
corresponding {\it tensorial} covariant differentiation operator 
$\nablab$ whose effect is given by
{\be \nablab_{\!\nu}\vb^\mu=\partial_\mu \vb^\mu+
\Gammab_{\!\mu \ \rho}^{\ \nu}\,\vb^\rho \, .\label{3.4}\fe}

In an exactly analogous manner, the connection on the fibre space will
be specified by another such connector field $\hat\Gammar$ with
components $\hat\Gammar{_{\!_\Ar\ _\Cr}^{\ _\Br}}$ whose use can be
illustrated as before by the case of a simple fibre-tangent vector,
$\Vr$ say, with components $\Vr^{_\Ar}$, whose covariant variation
$\hat\dr\Vr$ will be given in terms of corresponding component
variations ${\rm d}(\Vr^{_\Ar})$ and fibre displacement components
$\hat {\rm d}\Xr^{_\Ar}$ by
{\be (\hat\dr\Vr)^{_\Ar}= {\rm d} (\Vr^{_\Ar})+\hat
\Gammar{_{\!_\Br\ _\Cr}^{\ _\Ar}}\Vr^{_\Cr}{\rm d}\Xr^{_\Br}\label{3.5}\fe}
so that if we were concerned with a field defined over the fibre space
we would have a corresponding fibre-covariant differentiation operator
whose effect would be given by
{\be \hat\nablar_{\!_\Br} \Vr^{_\Ar}=\Vr^{_\Ar}_{\ ,_\Br}+\hat
\Gammar{_{\!_\Br\ _\Cr}^{\ _\Ar}}\Vr^{_\Cr}\, .\label{3.6}\fe}

What we are actually most interested in is situations where the entities
such as $\Vr$ under consideration are specified as fields not over the 
fibre space $\calXr$ but over the base space ${\cal M}$, or to be more 
explicit where they are specified as fields on some section $\Phir\{x\}$ 
of the bundle $\calBr$ with fibres $\calXr$ over ${\cal M}$. in such a 
situation we shall be concerned with variations for which the fibre
displacement ${\rm d}\Xr^{_\Ar}$ appearing in (\ref{3.5}) will be
determined (via the section $\Phir$) by a base-space displacement
${\rm d}x^\mu$ in the form
{\be {\rm d}\Xr^{_\Ar}= (\nablab_{\!\mu}\Phir^{_\Ar}){\rm d}x^\mu
\, ,\label{3.7}\fe}
where the {\it bitensorial gradient components} are defined by
{\be \nablab_{\!\mu}\Phir^{_\Ar}=\partial_\mu \Xr^{_\Ar}\{\Phir\{x\}\}
\, .\label{3.8}\fe}
There will thus be a corresponding {\it bitensorial} generalisation
of the covariant differentiation operator $\nablab$, whose
effect on a fibre-tangent field $\Vr$ at the section $\Phir$ over
${\cal M}$ will be given by
{\be \nablab_{\!\mu}\Vr^{_\Ar}=\partial_\mu \Vr^{_\Ar}+
\Gammar_{\!\mu \ _\Br}^{\ _\Ar}\Vr^{_\Br}\, ,\label{3.9}\fe}
where the (biaffinitorial) {\it section dependent} connector
components $\Gammar_{\!\mu \ _\Br}^{\ _\Ar}$ are given by 
{\be \Gammar_{\!\mu \ _\Br}^{\ _\Ar}=\Xr^{_\Cr}_{\ \mu}\,
\hat\Gammar{_{\!_\Cr\ _\Br}^{\ _\Ar}}\, \label{3.10}\fe}
using the abbreviation
{\be \Xr^{_\Cr}_{\ \mu}=\nablab_{\!\mu}\Phir^{_\Cr}\label{3.11}\fe}
for the components of the (gradient) projection bitensor defined
by the section $\Phir$ according to (\ref{3.8}).

Once the connectors $\Gammab_{\!\mu \ \rho}^{\ \nu}$ and 
$\Gammar_{\!\mu \ _\Br}^{\ _\Ar}$ are available, one can proceed at once
in the usual way to write down the covariant bitensorial derivatives
of bitensors of arbitrary orders by including a connector term of the
appropriate kind for each index. The lowest (zero) order example is
the case of the covariant derivative of the section $\Phir$ itself,
as given by (\ref{3.8}), for which no connector term is needed at all.

As one would expect, commuting the order of covariant differentiation
operations brings to light torsion and curvature effects resulting from
torsion and curvature in ${\cal M}$ and $\calXr$. The ordinary base-space
torsion and curvature are given by the usual expressions
{\be \Thetab_{\mu\nu}^{\,\ \ \rho}= 2\Gammab_{\![\mu \ \nu]}^{\, \ \rho} 
\label{3.12}\fe}
and 
{\be \Rb_{\mu\nu\ \sigma}^{\,\ \ \rho}=\partial_{[\mu}\Gammab_{\!\nu] \ \sigma}
^{\ \rho}+2\Gammab_{\![\mu \ |\tau|}^{\, \ \rho}\Gammab_{\!\nu]\ \sigma}
^{\  \tau} \, , \label{3.13}\fe}
while the analogous fibre torsion and curvature are defined similarly by
{\be \hat\Thetar{_{_{\Ar\Br}}^{\ \ \ _\Cr}}= 2\hat\Gammar{_{\![_\Ar \ _\Br]}
^{\, \ _\Cr}} \label{3.14}\fe}
and
{\be \hat\Rr{_{_{\Ar\Br}\ _\Dr}^{\ \ \ _\Cr}}=2\hat
\Gammar{_{\![_\Br\ |_\Dr| ,_\Ar]}^{\,\ _\Cr}}
+2\hat\Gammar{_{\![_\Ar \ |_\Er|}^{\, \ _\Cr}}\hat\Gammar{_{\!_\Br]\ _\Dr}
^{\  _\Er}} \, . \label{3.15}\fe}
In terms of these, the effect of commuting two covariant differentiations
at the zero level, i.e. when acting on the primary section $\Phir$ itself,
 will be given by
{\be 2\nablab_{\![\mu}\nablab_{\!\nu]}\Phir^{_\Ar}=\Xr^{_\Cr}_{\ \mu}
\Xr^{_\Dr}_{\ \nu}\,\hat\Thetar{_{_{\Cr\Dr}}^{\ \ \ _\Ar}}- 
\Thetab_{\mu\nu}^{\,\ \ \rho}\Xr^{_\Ar}_{\ \rho} \, .\label{3.16}\fe}
At the first order level, when acting on a base space vector field 
wa shall obtain an expression of the usual form
{\be 2\nablab_{\![\mu}\nablab_{\nu]}\vb^\rho=\Rb_{\mu\nu\ \sigma}^{\,\ \ \rho}
\vb^\sigma -\Thetab_{\mu\nu}^{\,\ \ \sigma}\nabla_{\!\sigma}\vb^\rho
\label{3.17} \fe}
and when acting on a fibre-tangent vector field we shall obtain
{\be 2\nablab_{\![\mu}\nablab_{\nu]}\Vr^{_\Ar}=\Rr_{\mu\nu\, \ _\Br}^{\ \ \ _\Ar}
\Vr^{_\Br}-\Thetab_{\mu\nu}^{\,\ \ \sigma}\nablab_{\!\sigma}\Vr^{_\Ar}
\label{3.18}\fe}
where the (bitensorial) {\it section-dependent} base projection of the
fibre curvature is given by
{\be\Rr_{\mu\nu\, \ _\Br}^{\ \ \ _\Ar}=\Xr^{_\Cr}_{\ \mu}\Xr^{_\Dr}_{\ \nu}\,
\hat\Rr{_{_{\Cr\Dr}\ _\Br}^{\ \ \ _\Ar}}\, .\label{3.19}\fe}

Having seen how the specification of the linear connections $\Gammab$ 
and $\hat\Gammar$ on the base and fibre spaces, ${\cal M}$ and $\calXr$ 
respectively, will automatically determine a natural bitensorial 
differentiation operator in the trivial case of a bundle with a 
direct-product structure (or equivalently with an integrable bundle 
connection) we now want to consider the case of the generalization of
this procedure to the case in which one has a nonintegrable bundle 
connection $\Amr$ in a bundle whose fibres are subject to a nontrivial
action of an automorphism group $\calGr$. As a preliminary to setting up
the actual gauge-covariant differentiation procedure in Sec. \ref{Sec5},
we shall first describe the appropriate {\it primary realization} of
the gauge algebra in terms of vertical fields on the primary bundle
$\calBr$.

\section{The primary fibre-tangent vector realization of a gauge field}
\label{Sec4}

Instead of supposing that the primary bundle has a preferred (or indeed 
any) direct-product structure (as was done in the previous section) we
now consider the more general situation in which the bundle fibres are
horizontally related only by a nonintegrable connection $\Ar$ subject
to a {\it nonintegrable} action of an automorphism group $\calGr$
with Lie algebra $\calAr$.

In this more general case, the bundle will still have a simple (albeit no
longer uniquely preferred) {\it local direct-product} strcture
$\calXr\times{\cal N}$, i.e. what is traditionally known as a {\it gauge},
above each (sufficiently small) neighbourhood ${\cal N}$ in the base
space ${\cal M}$: in terms of local coordinates $\Xr^{_\Ar}$ on some
local fibre-space patch $\calUr$ and $x^\mu$ on the base-space patch
${\cal N}$ the bundle points represented by the pair $(\Xr,x)$ with
$\Xr \in \calUr$, $x\in {\cal N}\subset {\cal M}$, will be specified by a
corresponding set of local gauge coordinates $\{\Xr^{_\Ar}, x^\mu\}$.
However \cite{1,4} it is now no longer required that any particular
such gauge (i.e. direct product) structure be preserved when the local
bundle patches are fitted together. Since a given gauge over ${\cal N}$
will specify an isomorphism mapping $\Jr\{x\}$ of the fibre over each
point $x\in {\cal N}$ into the abstract fibre space $\calXr$, and any
other gauge over an onerlapping neighborhood ${\cal N}^\prime$ will 
specify an analogous isomorphism $\Jr^\prime\{x\}$ for $x\in{\cal N}^\prime$,
it follows that there will be a corresponding isomorphism of the form
{\be\calXr\ \vbox{\hbox{$_{\,\Gr}$} \hbox{$\rightarrow$} } \ \calXr
\, ,\hskip 1.6 cm  \Xr \ \mapsto \ \Gr\Xr \label{4.1}\fe}
of the fibre space onto itself, determined for any $x\in {\cal N} \cap 
{\cal N}^\prime $ by the product mapping $\Gr=\Jr^\prime\circ \Jr^{-1}$.
If the second (new) gauge is represented in an overlapping patch
by the local gauge coordinates $\{\Yr^{_\Ar}, x^\mu\}$ where the $\Yr^{_\Ar}$
are coordinates on some local patch $\calUr\subset\calXr$, then there will 
be a relation of the general form (\ref{2.7}) specifying the {\it new}
gauge coordinates $\{\Gr^{_\Ar}, x^\mu\}$ of a point represented by the 
pair $(\Xr,x)$ with local coordinates $\{\Gr^{_\Ar}\{\Xr\}, x^\mu\{x\}\}$ 
in the {\it original} gauge by prescription of the form
{\be  \Gr^{_\Ar}\{\Xr,x\}=\Yr^{_\Ar}\{\Gr\{x\}\Xr\}\, .\label{4.2}\fe}

As usual the bundle connection over ${\cal M}$ will be determined by
the specification of a corresponding connector one-form
$\Amr_\mu$ with (gauge patch dependent) values in the Lie algebra
$\calAr$, and there will be a corresponding (gauge patch dependent) 
Lie algebra valued two-form
{\be \Fmr_{\mu\nu}= 2\partial_{[\mu}\Amr_{\nu]}+ 2 \Amr_{[\mu}\Amr_{\nu]}
\label{4.3}\fe}
satisfying a Bianchi identity of the form
{\be \partial_{[\mu}\Fmr_{\nu\rho]}+[\Amr_{[\mu},\Fmr_{\nu\rho]}]=0
\label{4.4}\fe}
and vanishing only if the connection is integrable.

In terms of a representation of the form
{\be \Amr_\mu=\Aru_\mu^{\ \alpharu}\, \amr_\alpharu \label{4.5}\fe}
in terms of a fixed basis $\amr_\alpharu\in \calAr$ ($\alpharu=$ 1, ... , m),
of the Lie algebra with structure constants specified by
{\be [\amr_\alpharu ,\amr_\betaru]=
\ConStruc_{\alpharu\betaru}^{\,\ \ \gammaru}\, \amr_\gammaru \label {4.6}\fe}
the corresponding curvature two-form components in the corresponding
representation
{\be \Fmr_{\mu\nu}=\Fru_{\!\mu\nu}^{\ \ \alpharu}\, \amr_\alpharu
\label{4.7}\fe}
have the explicit expression
{\be \Fru_{\!\mu\nu}^{\ \ \alpharu}=2\partial_{[\mu} \Aru_{\nu]}^{\ \alpharu}+
\ConStruc_{\betaru\gammaru}^{\,\ \ \alpharu}\,\Aru_\mu^{\ \betaru}
\Aru_\nu^{\ \gammaru}\, .\label{4.8}\fe}

\begin{figure}
\centering
\includegraphics[width=16cm]{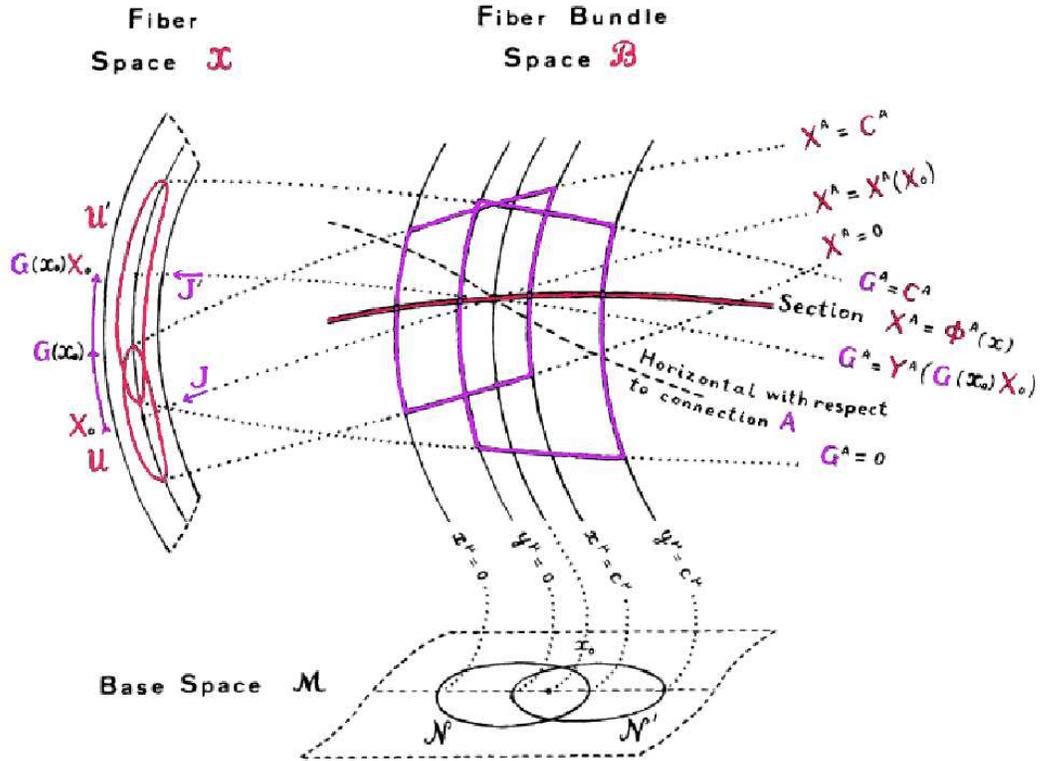}
\caption{Schematic representation showing two-dimensional subspaces of 
a (curved) fibre space $\calXr$, a base space ${\cal M}$, and a bundle
$\calBr$ with fibre $\calXr$ over ${\cal M}$, indicating the relationships
between the various local coordinate patches mentioned in the text,
and showing the distinction between the original gauge projection
$\Jr$ determined (for $\Xr\in \calXr$, $x\in {\cal M})$ in the form 
$\Jr\{\Xr,x\}=\Xr$ by the local product structure corresponding to some
initially given gauge over a neighbourhood ${\cal N}\subset {\cal M}$, and 
a {\it new} gauge projection $\Jr^\prime$ over ${\cal N}^\prime$ given in 
terms of the {\it initial} local product structure over the intersection
${\cal N} \cap {\cal N}^\prime$ by $\Jr^\prime\{\Xr,x\}=\Gr\{x\}\Xr$. 
(The positions of thepatches ${\cal N}$, ${\cal N}^\prime$, in ${\cal M}$
and  ${\cal U}$, ${\cal U}^\prime$, in ${\calXr}$ are indicated by pairs of
points representing the coordinate origin and some other arbitrary
constant values denoted by the letter c.)
.}
\label{fig1}
\end{figure}

In the simple vector bundles that are most commonly used in physics, 
the algebra $\calAr$ can conveniently be represented in terms of 
matrices, but in the general nonlinear case it is more useful to think 
of the algebra as represented by the {\it vector fields} that generate the
corresponding infinitesimal diffeomorphisms on the primary fibre space
$\calXr$ under consideration. The basic function of a gauge field
$\Amr$ is to determine, for any infinitesimal base displacement 
${\rm d}x$, a corresponding algebra element 
{\be \amr=\Amr_\mu\, {\rm d}x^\mu \label {4.9}\fe}
which will be realised by a corresponding fibre vector field with
components
{\be \aru^{_\Ar}=\Aru_\mu^{\ _\Ar} \, {\rm d}x^\mu \, .\label {4.10}\fe}
The role of this vector field is to determine the (infinitesimal)
deflection between the horizontal projection, as determined by the local
direct product structure associated with the gauge patch under 
consideration (in effect the local coordinates $\{\Xr, x\}$) between
 fibres over base points differing by the infinitesimal base 
displacement ${\rm d}x$, and the corresponding horizontal projection
as determined by the connection.

The specification of a connection in this way enables one to define
a gauge covariant vertical displacement $\dr\Xr$ between neighboring
points on neighboring fibres, as determined with respect to horizontality
as specified by the connection. The components of the covariant vertical
displacement may be evaluated as the difference,
{\be \dr\Xr^{_\Ar}= {\rm d}\Xr^{_\Ar}-{\rm d}_\amr \Xr^{_\Ar}
\label{4.11}\fe}
between the vertical deviation ${\rm d}\Xr^{_\Ar}$ determined by
the local coordinates (i.e. by the local product structure of the
gauge path) and the vertical deviation
{\be {\rm d}_\amr \Xr^{_\Ar}= -\aru^{_\Ar} \label{4.12}\fe}
between horizontality with respect to the connection and horizontality
as determined by the local coordinates. Hence if we are considering a 
section  $\Phir$, substitution of the corresponding coordinate 
displacement formula
{\be  {\rm d}\Xr^{_\Ar}= \Xr^{_\Ar}_{\ \mu}\, {\rm d}x^\mu \label{4.13}\fe}
into (\ref{4.11}) gives the expression

{\be  \dr\Xr^{_\Ar}=(\Xr^{_\Ar}_{\ \mu} +\Aru^{_\Ar}_{\ \mu}) 
\, {\rm d}x^\mu \label{4.14}\fe}
for the corresponding covariant displacement components, where 
$\Xr^{_\Ar}_{\ \mu}$ are the tangent projection components associated with 
the section $\Phir$ as given by (\ref{3.11}). (See Fig. \ref{fig1}.)

It is evident that the quantity $\dr\Xr^{_\Ar}$ constructed in this
way will be vectorially covariant under the effect of a {\it general}
(base-position dependent) fibre-coordinate transformation of the form
(\ref{2.7}) which gives
 {\be {\rm d}\Xr^{_\Ar}\ \mapsto \ \Gr^{_\Ar}_{\  _\Br}\,  {\rm d}\Xr^{_\Br}
+ \Gr^{_\Ar}_{\ ,\mu} \, {\rm d} x^\mu \label{4.15}\fe}
provided that the gauge connection field $\Amr$ undergoes the
corresponding transformation, which will be given explicitly for 
the vector realization by
{\be \Aru_\mu^{\ _\Ar}\ \mapsto\ \Gr^{_\Ar}_{\  _\Br}\Aru_\mu^{\ _\Br}
- \Gr^{_\Ar}_{\ ,\mu} \label{4.16}\fe}
since the inhomogeneous terms will cancel so as to give the purely
vectorial covariance rule
{\be  \dr\Xr^{_\Ar}\ \mapsto \Gr^{_\Ar}_{\  _\Br}\, \dr\Xr^{_\Br}\, .
\label{4.17}\fe}

By a rather longer calculation one can also verify that (\ref{2.7}) and
(\ref{4.16}) also imply an analogous purely vectorial covariance rule
{\be \Fru_{\!\mu\nu}^{\ \ _\Ar}\ \mapsto \Gr^{_\Ar}_{\  _\Br}\,
 \Fru_{\!\mu\nu}^{\ \ _\Br} \label{4.18}\fe}
for the components of the vector realisation of the gauge curvature
$\Fmr$, as defined by
{\be \Fru_{\!\mu\nu}^{\ \ _\Ar}=\Fru_{\!\mu\nu}^{\ \ \alpharu}\,
\aru_\alpharu^{\, _\Ar} \, \label{4.19}\fe}
where $\aru_\alpharu^{\, _\Ar}$ are the components of the vector realization
of $\amr_\alpharu$, and the basis components $\Fru_{\!\mu\nu}^{\ \ \alpharu}$ 
of the gauge curvature are specified by (\ref{4.8}).

Since the algebra commutator relations will be realised by the Lie
differentiation commutator of the vector fields on $\calXr$, the structure
relations (\ref{4.6}) will be realised concretely by 
{\be 2 \aru_{[\alpharu|\, ,_\Br}^{\ \ _\Ar}\aru_{|\betaru]}^{\,\ _\Br}=
\ConStruc_{\alpharu\betaru}^{\,\ \ \gammaru}\, \aru_\gammaru^{\, _\Ar}
\, .\label{4.20}\fe}
Hence by substitution in (\ref{4.8}) we obtain an explicit, 
Lie algebra-basis independent, expression for the components
$ \Fru_{\!\mu\nu}^{\ \ _\Ar}$ of the realization of the gauge curvature
$\Fmr$, namely
{\be\Fru_{\!\mu\nu}^{\ \ _\Ar}= 2 \Aru_{[\nu\,\ ,\,\mu]}^{\ \ _\Ar}+ 2
\Aru_{[\nu}^{\, \ _\Br} \Aru_{\mu] \ ,_\Br}^{\, \ _\Ar}\, .\label{4.21}\fe}

It is an essentially straightforward exercise in partial differentiation 
to verify directly that this {\it fundamental primary bundle realisation}
of the gauge curvature does indeed undergo a transformation of the 
vectorial form (\ref{4.18}) under the effect of a general gauge-patch 
transformation as specified by (\ref{2.7}) and (\ref{4.17}). This
establishes that the base-space two-form valued vertical (i.e. fibre-space
tangent) vector field $\Fmr$ specified by (\ref{4.18}) is {\it globally
well defined} over the whole of the primary bundle $\calBr$, unlike
the base space one-form valued vertical vector field $\Amr$ which
is gauge patch dependent.

The property of existing as a field over the whole of the primary bundle
$\calBr$ distinguishes the primary gauge-curvature realisation $\Fmr$
from the other bitensorial entities introduced in the previous sections,
which were defined only over some particular section $\Phir$ in $\calBr$.
In dealing with entities such as $\Fmr$ and $\Amr$ which are defined over
the whole of the fibres and not just at the section $\Phir$, one must
take care to distinguish the partial component derivatives, indicated by
a comma, from the total base-space gradient components for the field over
${\cal M}$ that would be determined by the section $\Phir$. Thus 
although we could use the expressions $\partial_\mu\Aru_\nu^{\ \alpharu}$ and
$\Aru_{\nu\ ,\,\mu}^{\ \alpharu}$ interchangeably in (\ref{4.8}), it is
important to notice that $\partial_\mu\Aru_\nu^{\ _\Ar}$ is not the
same as $\Aru_{\nu\ ,\, \mu}^{\ _\Ar}$ in the algebra-basis independent 
expression (\ref{4.21}), the distinction being specified as a function 
of the section $\Phir$ by
{\be \partial_\mu\Aru_\nu^{\ _\Ar}=\Aru_{\nu\ ,\, \mu}^{\ _\Ar}+
\Aru_{\nu\ , _\Br}^{\ _\Ar} \Xr^{_\Br}_{\ \mu}\, . \label{4.22}\fe}
By paying attention to this distinction, it will be possible to work
with an explicit, but Lie algebra-basis independent notation scheme
throughout the remainder of this work, thereby avoiding any further
reference to such cumbersome paraphenalia as the structure constants.

Up to this point we have made no reference to any specific properties
of the gauge group $\calGr$: the analysis in this section
would be valid for transformations $\Xr^{_\Ar} \mapsto \Gr^{_\Ar}$
resulting from the general action of the {\it entire (infinite parameter)
group of diffeomorphisms} on the fibre space $\calXr$. However, for the
purpose of constructing a gauge covariant differentiation formalism, as
will be done in the section that follows, it will be necessary to restrict
ourselves to situations for which $\calGr$ is included in the at most
finite-dimensional diffeomorphism subgroup leaving the chosen fibre-space
connection $\hat\Gammar$ invariant.

\section{Gauge-covariant bitensorial differentiation}
\label{Sec5}

It is immediately evident from the work of the previous section that for
any section $\Phir\{x\}$ in the primary bundle $\calBr$ the gauge
connection $\Amr$ will determine a well defined covariant vector field
$\Dru\Phi$ over the base-space ${\cal M}$, whose components can be read 
out from the expression
{\be \dr\Xr^{_\Ar}=\Phir^{_\Ar}_{\ \wru\,\mu} \, {\rm d}x^\mu
\label{5.1}\fe}
for the covariant vertical displacement $\dr\Xr$ resulting from a
base-space displacement ${\rm d}x$, where we have intrioduced a heavy
{\it bar} notation convention
{\be \Dru_\mu\Phir^{_\Ar}=\Phir^{_\Ar}_{\ \wru\,\mu} \label{5.2}\fe}
for gauge-covariant differentiation. Recalling our previous abbreviation
{\be \partial_\mu\Phir^{_\Ar}=\Xr^{_\Ar}_{\ \mu} \label {5.3}\fe}
we immediately obtain the compact expression
{\be \Phir^{_\Ar}_{\ \wru\,\mu}=\Xr^{_\Ar}_{\ \mu}+\Aru_\mu^{\ _\Ar}
\label{5.4}\fe}
for the bitensorial derivative components $ \Phir^{_\Ar}_{\ \wru\,\mu}$ by
substituting (\ref{4.14})  in (\ref{5.1}).

This lowest order differentiation procedure obviously does not depend 
on the specification of any intrinsic structure on the fibre $\calXr$ or 
base ${\cal M}$ of $\calBr$. However, in order to go on (analogously to 
the work of Sec. \ref{Sec3}) to the construction of higher order 
bitensorial derivatives, the reintroduction of the fibre connection
$\hat\Gammar$ on $\calXr$ and, more routinely, of the base connection
$\Gammab$ on ${\cal M}$, will evidently be necessary.

Before continuing, we now make the usual supposition that the gauge 
group $\calGr$ acting effectively on the primary bundle $\calBr$ should
be restricted so as to consist only of fibre isomorphisms, i.e. that it
should leave invariant all relevant structure on the fibre space $\calXr$ 
in which the primary field is evaluated. As a minimal requirement we must 
at least demand that the transformation group $\calGr$ should preserve
the only structure that has been introduced so far on $\calXr$, namely
the indispensible fibre connection $\hat\Gammar$; i.e. the gauge 
transformations must be restricted so as not to violate the essential 
property
{\be \hat\Gammar{_{\!_\Ar\ _\Cr\, ,\, \mu}^{\ _\Br}}=0 \label{5.5}\fe}
characterising any allowable local gauge coordinate system
$\{\Xr,x\}$. In order to express the corresponding restiction on the
gauge algebra, it is convenient, following Yano\cite{5}, to introduce
an abbreviation, which we shall indicate by a subscript colon, to 
indicate a covariant derivative of a vector field that differs from the
usual one in that the connection is inserted the {\it wrong way round}.
Thus for the particular case of the gauge vector one form $\Amr$ we
introduce a corresponding gauge tensor one form $\Amr_{:}$ defined by
{\be \Aru_{\mu\ :_\Br}^{\ _\Ar}= \Aru_{\mu\ ,_\Br}^{\ _\Ar}
+\Aru_\mu^{\ _\Cr}\, \hat\Gammar{_{\!_\Cr\ _\Br}^{\ _\Ar}} \label{5.6}\fe}
or equivalently
{\be \Aru_{\mu\ :_\Br}^{\ _\Ar}= \hat\nablar_{\!_\Br}\Aru_{\mu}^{\ _\Ar}
+\Aru_\mu^{\ _\Cr}\, \hat\Thetar{_{\!_{\Cr\Br}}^{\ \ _\Ar}} \, ,\label{5.7}\fe}
where $\hat\nablar$ denotes the {\it ordinary} operation of covariant
differentiation along the fibres with respect to the fibre connection
$\hat\Gammar$. In the absence of the torsion $\hat\Thetar$ the distinction
between this Yano covariant derivative and the ordinary covariant 
derivative disappears. In terms of this notation, the essential requirement 
that the fibre connection be invariant under the action generated by the
primary gauge field $\Amr$ can be obtained (from Yano's formula \cite{5} for
the Lie derivative of the connection) in the form
{\be  \hat\nablar_{\!_\Br}\Aru_{\mu\ :_\Cr}^{\ _\Ar}=\Aru_\mu^{\ _\Dr}\,
\hat \Rr{_{_{\Br\Dr}\ _\Cr}^{\ \ \ _\Ar}}
\label{5.8}\fe}
This basic postulate includes, as a consequence, the corresponding
decoupled invariance requirement for the torsion tensor, i.e.
{\be \Aru_\mu^{\ _\Dr}\hat\nablar_{\!_\Dr} 
\hat\Thetar{_{\!_{\Br\Cr}}^{\ \ _\Ar}} = \Aru_{\mu\ :_\Dr}^{\ _\Ar}
\hat\Thetar{_{\!_{\Br\Cr}}^{\,\ \ _\Dr}}+2\Aru_{\mu\ :[_\Cr|}^{\ _\Dr}
\hat\Thetar{_{\!_{\Dr}|_\Br]}^{\,\ \ \ _\Ar}}\, .\label{5.9}\fe}

For purely base-tensorial entities  the question of gauge invariance does 
not arise. We therefor proceed directly to consider the appropriate
gauge-covariant generalization of the definition (\ref{3.5}) of the 
absolute variation of the simplest kind of fibre-tensorial quantity, an 
ordinary vector $\Vr$, between nearby points in nearby fibres separated by
a base displacement ${\rm d}x$. Evidently the required gauge-covariant
variation $\dr\Vr$ should be defined as the covariant variation with
respect to the fibre connection $\hat\Gammar$ along the vertical 
displacement $\dr\Xr$ as specified by the projection that is horizontal
with respect to the gauge connection. This means that we must take
{\be \dr\Vr^{_\Ar}={\rm d}(\Vr^{_\Ar})- {\rm d}_{\amr}\Vr^{_\Ar}+
(\dr\Xr^{_\Br})\,\hat\Gammar{_{\!_\Br\ _\Cr}^{\ _\Ar}} \Vr^{_\Cr} \, ,
\label{5.10}\fe}
where $\dr\Xr^{_\Ar}$ are the components of the covariant veritical 
displacement as specified by (\ref{4.11}), or more explicitely 
(\ref{4.14}), and ${\rm d}_{\amr}\Vr^{_\Ar}$ are the vector component 
variations  resulting from the fact that horizontality with respect to
the local fibre coordinates $\Xr^{_\Ar}$ differs from horizontality with
respect to the gauge connection by the effect of the infinitesimal Lie
displacement induced by the vector field $\amr$ specified on the fibre
by (\ref{4.10}), which gives 
 {\be {\rm d}_{\amr}\Vr^{_\Ar} = -\aru^{_\Ar}_{\ ,_\Br}\Vr^{_\Br}\, .
\label{5.11}\fe}
Thus explicitly we shall have 
{\be \dr\Vr^{_\Ar}={\rm d}(\Vr^{_\Ar}) +\left[ \Aru_{\mu\ ,_\Br}^{\ _\Ar}
{\rm d} x^\mu +(\Aru_\mu^{\ _\Cr}\, {\rm d}x^\mu+ {\rm d}\Xr^{_\Cr})
\hat\Gammar{_{\!_\Cr\ _\Br}^{\ _\Ar}}\right] \Vr^{_\Br}\, .
\label{5.12}\fe}

In the case where $\Vr$ is a tangent vector defined as a field on a section
$\Phir\{x\}$, there will be a corresponding bitensorial covariant 
derivative which can be read out from the defining formula
{\be \dr\Vr^{_\Ar}=\Vr^{_\Ar}_{\ \wru\,\mu}\,  {\rm d}x^\mu \label{5.13}\fe}
using the abbreviated bar suffix notation system
{\be\Dru_\mu\Vr^{_\Ar}=\Vr^{_\Ar}_{\ \wru\,\mu}\, . \label{5.14}\fe}
Thus we obtain the covariant derivative components in the form
{\be \Vr^{_\Ar}_{\ \wru\,\mu}=\partial_\mu\Vr^{_\Ar}+
\omegaru_{\mu \ _\Br}^{\ _\Ar}\Vr^{_\Br}\, ,\label{5.15}\fe}
where the section dependent {\it connector} $\omegaru$ [as introduced in 
(\ref{2.4})] will be given, using the notation of (\ref{5.4}), by
{\be \omegaru_{\mu \ _\Br}^{\ _\Ar}=\Phir^{_\Cr}_{\ \wru\,\mu} \hat
\Gammar{_{\!_\Cr\ _\Br}^{\ _\Ar}}+  \Aru_{\mu\ ,_\Br}^{\ _\Ar} \label{5.16}\fe}
or equivalently, using the notation of (\ref{3.10}) and (\ref{5.6}),
 {\be \omegaru_{\mu \ _\Br}^{\ _\Ar}= \Gammar_{\!\mu\ _\Br}^{\ _\Ar} +
\Aru_{\mu\ :_\Br}^{\ _\Ar} \, . \label{5.17}\fe}

Having thus obtained the required connector $\omegaru$ that is needed for
covariant differentiation of a simple fibre vector on the section, one 
can go on immediately in the usual way to construct the corresponding 
covariant derivatives of more general fibre-tensorial and bitensorial
quantities by adding an appropriate connector term for each index (a term 
involving $\omegaru_{\mu \ _\Br}^{\ _\Ar}$ with a positive or negative sign 
for each respectively contravariant or covariant fibre index,and 
similarly a term involving $\Gammab_{\!\mu \ \rho}^{\ \nu}$ for each
base-space index.

The resulting generalisation of the derivative commutator rule (\ref{3.16})
for the primary section $\Phir$ itself involves the fibre and base torsions 
and the gauge curvature, taking the form
{\be 2\Phir^{_\Ar}_{\ \wru\,[\nu\,\wru\,\mu]}=\Phir^{_\Cr}_{\ \wru\,\mu}
\Phir^{_\Dr}_{\ \wru\,\nu}\,\hat\Thetar{_{_{\Cr\Dr}}^{\ \ \ _\Ar}} 
-\Thetab_{\mu\nu}^{\,\ \ \rho}\,\Phir^{_\Ar}_{\ \wru\,\rho}+
\Fru_{\!\mu\nu}^{\ \ _\Ar} \, . \label{5.18}\fe}

The analogous commutator rule, generalising (\ref{3.18}), for a fibre
vector field over the section $\Phir$ involves the fibre curvature and
the gauge curvature as well as the base torsion, taking the form
{\be 2\Vr^{_\Ar}_{\ \wru\,[\nu\,\wru\,\mu]}=\Omegaru_{\mu\nu \ _\Br}
^{\,\ \ _\Ar}\Vr^{_\Br}-\Thetab_{\mu\nu}^{\,\ \ \rho}\,
\Vr^{_\Ar}_{\ \wru\,\rho} \, , \label{5.19}\fe}
where the total curvature, as defined by (\ref{2.11}), can be evaluated,
(using (\ref{5.8}) and (\ref{4.21}), as the sum of two {\it separately} 
bitensorially covariant terms, in the form
{\be \Omegaru_{\mu\nu \ _\Br}^{\,\ \ _\Ar}=\Phir^{_\Cr}_{\ \wru\,\mu}
\Phir^{_\Dr}_{\ \wru\,\nu}\, \hat\Rr{_{_{\Cr\Dr}\ _\Br}^{\ \ \ _\Ar}}+
\Fru_{\!\mu\nu\, \ :_\Br}^{\  \ _\Ar} \, . \label{5.20}\fe}
The first (section dependent) gauge covariant term on the right-hand 
side in (\ref{5.20}) can evidently be expanded quadratically in the
gauge connection field as
{\be\Phir^{_\Cr}_{\ \wru\,\mu}\Phir^{_\Dr}_{\ \wru\,\nu}
\, \hat\Rr{_{_{\Cr\Dr}\ _\Br}^{\ \ \ _\Ar}}= 
\Rr_{\mu\nu\ _\Br}^{\ \ \ _\Ar}+ 2\Aru_{[\mu}^{\,\ _\Cr}\Xr^{_\Dr}_{\ \nu]} 
\hat \Rr{_{_{\Cr\Dr}\ _\Br}^{\ \ \ _\Ar}}+\Aru_{\mu}^{\ _\Cr}
\Aru_{\nu}^{\ _\Dr}\hat \Rr{_{_{\Cr\Dr}\ _\Br}^{\ \ \ _\Ar}}
\, . \label{5.21}\fe}
We recapitulate that in the second term on the right-hand side in 
(\ref{5.20}) the colon denotes the Yano type (wrong way round) covariant
derivative, i.e.
{\be \Fru_{\!\mu\nu\, \ :_\Br}^{\  \ _\Ar}=\hat\nablar_{\!_\Br}
 \Fru_{\!\mu\nu}^{\  \ _\Ar} + \Fru_{\!\mu\nu}^{\  \ _\Cr}\, \hat
\Thetar{_{_{\Cr\Br}}^{\ \ _\Ar}}\, . \label{5.22}\fe}
Like the undifferentiated curvature field $\Fmr$ itself, this Yano
gauge-curvature gradient is well defined globally over the primary bundle 
(not just on the section $\Phir$ where $\omegaru$ and  $\Omegaru$ are 
defined). Since the gauge curvature belongs, by construction, to the Lie 
algebra it will automatically satisfy a fibre-connection preservation 
condition of a form analogous to the fundamental requirement (\ref{5.8})
namely 
{\be \hat\nablar_{\!\Br} \Fru_{\!\mu\nu\, \ :_\Cr}^{\  \ _\Ar}= 
\Fru_{\!\mu\nu}^{\ \ _\Dr}\, \hat \Rr{_{_{\Br\Dr}\ _\Br}^{\ \ \ _\Cr}}
\, .\label{5.23}\fe}
This relation is useful for the purpose of verifying directly as an 
exercise that the section-dependent curvature $\Omegaru$  given by
(\ref{5.20}) does indeed satisfy the Bianchi identity (\ref{2.12}).

\bigskip
{\bf Acknowledgements}
\medskip

 The author wishes to thank D. Bernard and N. Sanchez  for comments and
suggestions.

\end{document}